\documentclass[
]{ceurart}

\usepackage{listings}
\usepackage{rotating}
\begin{document}

\copyrightyear{2025}
\copyrightclause{Copyright for this paper by its authors. Use permitted under Creative Commons License Attribution 4.0 International (CC BY 4.0).}
\conference{5th International Workshop on Scientific Knowledge: Representation, Discovery, and Assessment, Nov 2024, Nara, Japan}

\title{AI4DiTraRe: Building the BFO-Compliant Chemotion Knowledge Graph}

\author[1,2]{Ebrahim Norouzi}[%
orcid=0000-0002-2691-6995,
email=Ebrahim.Norouzi@fiz-Karlsruhe.de,
]
\cormark[1]
\author[3]{Nicole Jung}[%
orcid=0000-0001-9513-2468,
email=nicole.jung@kit.edu,
]
\author[1]{Anna M. Jacyszyn}[%
orcid=0000-0002-5649-536X,
email=Anna.Jacyszyn@fiz-Karlsruhe.de,
]
\author[1]{Jörg Waitelonis}[%
orcid=0000-0001-7192-7143,
email=joerg.waitelonis@fiz-karlsruhe.de,
]
\author[1,2]{Harald Sack}[%
orcid=0000-0001-7069-9804,
email=Harald.Sack@fiz-Karlsruhe.de,
]

\address[1]{FIZ Karlsruhe -- Leibniz Institute for Information Infrastructure, Hermann-von-Helmholtz-Platz 1, 76344 Eggenstein-Leopoldshafen, Germany}
\address[2]{Karlsruhe Institute of Technology, Institute of Applied Informatics and Formal Description Methods, Kaiserstr. 89, 76133 Karlsruhe}
\address[3]{Karlsruhe Institute of Technology, Institute of Biological and Chemical Systems,
Hermann-von-Helmholtz-Platz 1, 76344 Eggenstein-Leopoldshafen, Germany}
\cortext[1]{Corresponding author.}
%\fntext[1]{These authors contributed equally.}

\begin{abstract}
Chemistry is an example of a discipline where the advancements of technology have led to multi-level and often tangled and tricky processes ongoing in the lab. The repeatedly complex workflows are combined with information from chemical structures, which are essential to understand the scientific process. An important tool for many chemists is Chemotion, which consists of an electronic lab notebook and a repository. This paper introduces a semantic pipeline for constructing the BFO-compliant Chemotion Knowledge Graph, providing an integrated, ontology-driven representation of chemical research data. The Chemotion-KG has been developed to adhere to the FAIR (Findable, Accessible, Interoperable, Reusable) principles and to support AI-driven discovery and reasoning in chemistry. Experimental metadata were harvested from the Chemotion API in JSON-LD format. The JSON-LD, as an RDF serialization, was ingested into the triple store and subsequently transformed into a Basic Formal Ontology-aligned graph through SPARQL CONSTRUCT queries. The source code and datasets are publicly available via GitHub. The Chemotion Knowledge Graph is hosted by FIZ Karlsruhe Information Service Engineering. Outcomes presented in this work were achieved within the Leibniz Science Campus ``Digital Transformation of Research'' (DiTraRe) and are part of an ongoing interdisciplinary collaboration.
%abstract should be plain text - links are moved to introduction
%\url{https://github.com/ISE-FIZKarlsruhe/chemotion-kg}. The Chemotion Knowledge Graph is hosted at \url{https://ditrare.ise.fiz-karlsruhe.de/chemotion-kg/}.
\end{abstract}

\begin{keywords}
  digitalisation \sep
  chemistry \sep
  ontology \sep
  knowledge graph
\end{keywords}

\maketitle

\section{Introduction}
\label{sec:Introduction}

The generation of FAIR data relies on the ability to easily apply standards and to produce well-structured, well-annotated datasets. Electronic Lab Notebooks (ELNs) are essential tools in promoting the digitalization of scientific research, as they help incorporate standards and structure into the workflows of experimental scientists \cite{steinbeck2020nfdi4chem}.

In the field of chemistry, the development of an ELN that supports the creation of findable, accessible, interoperable, and reusable (FAIR) data is particularly challenging. This complexity arises from the nature of chemical research, which involves intricate and highly diverse experimental workflows, the use of various measurement devices that generate large volumes of data, and a wide range of data formats. Additionally, the handling of chemical structures which must be drawn, interpreted, and processed is indispensable for documenting and understanding chemical experiments. These requirements among others make chemistry data management especially demanding compared to other disciplines.

In recent years, a team of software developers and scientists at the Karlsruhe Institute of Technology (KIT) has been working on a research data management (RDM) environment specifically designed for chemistry. This environment, called Chemotion, comprises an ELN \cite{tremouilhac2017chemotion,Kotov2018chemotion2} tailored to the specific needs of chemists, along with a research data repository \cite{Tremouilhac2020chemotionrepo} that interoperates seamlessly with the ELN. The repository allows researchers to publish their recorded and analyzed data in accordance with FAIR principles.

The combination of the Chemotion ELN and the interoperable repository Chemotion provides scientists with a comprehensive digital infrastructure that addresses their domain-specific requirements. It enables them to manage, organize, analyze, and publish their data in a way that enhances transparency and ensures the reusability of their research outcomes \cite{Tremouilhac2021chemotionrepo,Huang2025chemotionrepo,Huang2021chemspectra}.

Driven by the needs of scientists for improved documentation and data management, particularly in response to the unique challenges posed by chemistry data, the Chemotion systems were initially developed without a systematic approach for integrating ontologies and semantic meaning. This limitation is now being addressed incrementally, through the implementation of a semantically coherent framework capable of representing research lifecycle stages, agent roles, provenance information, and domain-specific chemical entities in a machine-interpretable manner.

%this paragraph is moved to chapter "Impact"
%The overarching goal is to prepare the Chemotion system to fully leverage the potential of AI-based support and to enable its integration into the broader ecosystem of semantically enriched RDM systems. As part of this effort, artificial intelligence methods, including natural language processing, are being explored to automate data curation, particularly in the context of reaction descriptions and analysis modules within the Chemotion repository.8, 9 The development of fully automated curation workflows is expected to significantly enhance data quality and streamline system performance. Ultimately, these advancements will contribute to the realization of AI-assisted chemistry and support the emergence of self-driving laboratories10, which rely heavily on the automated analysis of chemical data.

To support and enhance the semantic enrichment of Chemotion, an ontology-driven modeling approach grounds the Chemotion Knowledge Graph (Chemotion-KG) in upper-level semantics provided by the Basic Formal Ontology (BFO) using the Chemotion repository as a source as it includes well curated data assigned to publication metadata. Ontology Design Patterns (ODPs) are adopted as a systematic method to encode these structures, serving as reusable solutions to recurring ontology engineering problems \cite{gangemi2009ontologydesignpatter}. Their use ensures not only the semantic consistency of datasets, creators, studies, and chemical substances in Chemotion-KG but also the reuse of established modeling practices that facilitate alignment with broader scientific knowledge graphs and ontology standards. 

The Chemotion-KG, a synergy of Chemotion and semantics, is being created within the Leibniz Science Campus ``Digital Transformation of Research'' (DiTraRe)\footnote{DiTraRe web page, \url{https://www.ditrare.de/en}} \cite{ditrare_proposal,jacyszyn2025ditrare_interim}. DiTraRe studies effects of digitalisation of research in a multilevel, interdisciplinary way. In this multidimensional exchange, one of the use cases is Chemotion which represents novel methods of data acquisition. The DiTraRe use cases are being analysed from different aspects by teams called {\it dimensions}. ``Exploration and Knowledge Organisation'' (AI4DiTraRe) is the dimension responsible for applying AI and studying its effects \cite{jacyszyn2024ditrare}. Collaboration of the aforementioned dimension and the Chemotion use case has by now provided outcomes described in this paper.

In brief, this work presents the construction of the BFO-compliant Chemotion-KG for semantically integrating experimental chemistry data. Section~\ref{sec:methodology} describes the workflow for data harvesting, semantic transformation, and ontology alignment. Section~\ref{sec:results} presents the resulting knowledge graph structures and instantiated entities generated using ontology design patterns. Section~\ref{sec:impact} discusses the broader impact of the Chemotion-KG and future directions. Finally, section~\ref{sec:Summary} provides a summary of the achieved contributions.

The source code and datasets are publicly available\footnote{\url{https://github.com/ISE-FIZKarlsruhe/chemotion-kg}}. The Chemotion-KG is hosted at \url{https://ditrare.ise.fiz-karlsruhe.de/chemotion-kg/}.

\section{Knowledge Graph Construction Approach}
\label{sec:methodology}

\begin{figure}[ht]
    \centering
    \includegraphics[width=\textwidth]{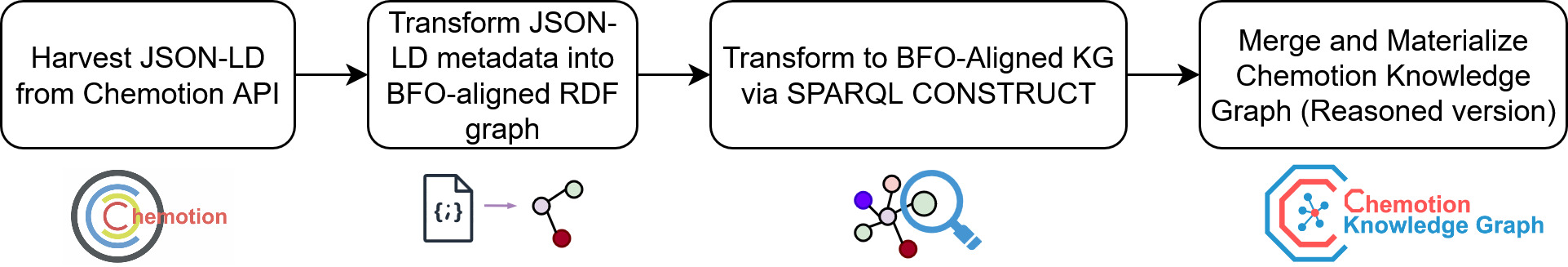}
    \caption{Schematic workflow of the Chemotion Knowledge Graph construction.}
    \label{fig:chemotion-kg-approach}
\end{figure}

This section describes the end-to-end pipeline for constructing the BFO-compliant Chemotion-KG, covering metadata harvesting, semantic enrichment, ontology alignment, reasoning, and materialization. The workflow is depicted in Figure~\ref{fig:chemotion-kg-approach}, illustrating the transformation from schema-based metadata to a semantically aligned knowledge graph.

The initial metadata were harvested from the Chemotion repository\footnote{\url{https://www.chemotion-repository.net}}, which provides structured schema.org-based descriptions of chemical research data. The repository's data covered (in)organic reactions, analytical measurements such as nuclear magnetic resonance (NMR) spectroscopy, mass spectrometry (MS), IR-, and Raman spectroscopy data. The metadata were expressed primarily with \texttt{schema:Dataset}, \texttt{schema:ChemicalSubstance}, and \texttt{schema:Study}, which provide only lightweight descriptive semantics. Consequently, while adequate for basic metadata exchange, the schema.org representation did not capture the rich semantic structure required for ontology-driven reasoning, provenance modeling, and alignment with upper-level ontologies such as the BFO\footnote{\url{https://basic-formal-ontology.org/}}\cite{bfo2015}.

To address this, the harvested JSON-LD was converted into RDF graphs with canonical URIs and semantically enriched through a SPARQL CONSTRUCT transformation. The mapping aligned the schema-based metadata to the BFO, leveraging the NFDICore ontology\footnote{\url{https://ise-fizkarlsruhe.github.io/nfdicore/}}\cite{bruns2024nfdicore} for research data structures, lifecycle modeling, and provenance, and reusing ChEBI\footnote{\url{https://www.ebi.ac.uk/chebi/}}\cite{CHAMEO} for chemical entities. The transformation queries were authored and documented using \texttt{shmarql}\footnote{\url{https://github.com/epoz/shmarql}}, a Linked Data publishing platform supporting SPARQL-based data pipelines. 

The overall SPARQL CONSTRUCT query implementing the semantic transformation is openly available in our GitHub repository\footnote{\url{https://github.com/ISE-FIZKarlsruhe/chemotion-kg/blob/main/processing/all-nfdicore.py}}. An example SPARQL UPDATE used to semantically enrich a dataset is shown below. It replaces a plain \texttt{schema:Dataset} description with NFDICore-aligned metadata, preserving original values while adding explicit BFO-compliant types. One of the essential steps in this transformation is the use of \texttt{BIND} to create new IRIs for elements such as descriptions, identifiers, titles, and URLs. Instead of representing these values only as plain literals, they are modeled as separate instances (e.g., description node, identifier node) with their own URIs. This approach follows BFO principles, where information content entities are treated as individual resources that can carry provenance and lifecycle information.

Creating explicit IRIs for these nodes allows the knowledge graph to keep stable references to metadata entities even when their literal values change. For instance, if a dataset's description or identifier is updated, the corresponding instance remains the same, making it possible to track changes over time and preserve provenance.

\begin{lstlisting}[language=SPARQL,caption={SPARQL CONSTRUCT for dataset transformation.}]
PREFIX schema: <http://schema.org/>
PREFIX nfdicore: <https://nfdi.fiz-karlsruhe.de/ontology/>
PREFIX obo: <http://purl.obolibrary.org/obo/>

CONSTRUCT {
  # Recast the dataset to BFO-aligned NFDICore class
  ?dataset a nfdicore:NFDI_0000009 ;               # Dataset
           nfdicore:NFDI_0001027 ?creator ;        # has creator (Person)
           nfdicore:NFDI_0000191 ?publisher ;      # has publisher (Organization)
           obo:IAO_0000235 ?descriptionNode ;      # description node
           nfdicore:NFDI_0001006 ?identifierNode ; # identifier node
           nfdicore:NFDI_0000142 ?license ;        # license information
           nfdicore:NFDI_0000216 ?technique ;      # associated measurement technique
           obo:IAO_0000235 ?nameNode ;             # title node
           obo:IAO_0000235 ?urlNode ;              # landing page URL node
           nfdicore:NFDI_0001023 ?study ;          # is output of a study
           obo:BFO_0000178 ?catalog .              # is part of a registered catalog
}
WHERE {
  # Original Chemotion metadata described using schema.org
  ?dataset a schema:Dataset ;
           schema:creator ?creator ;
           schema:publisher ?publisher ;
           schema:description ?description ;
           schema:identifier ?identifier ;
           schema:license ?license ;
           schema:measurementTechnique ?technique ;
           schema:name ?name ;
           schema:url ?url ;
           schema:includedInDataCatalog ?catalog ;
           schema:isPartOf ?study .

  # Generate canonical URIs for literal-based nodes to ensure global identification
  BIND(IRI(CONCAT("https://ditrare.ise.fiz-karlsruhe.de/chemotion-kg/nodes/", 
                 ENCODE_FOR_URI(?description))) AS ?descriptionNode)
  BIND(IRI(CONCAT("https://ditrare.ise.fiz-karlsruhe.de/chemotion-kg/nodes/", 
                 ENCODE_FOR_URI(?identifier))) AS ?identifierNode)
  BIND(IRI(CONCAT("https://ditrare.ise.fiz-karlsruhe.de/chemotion-kg/nodes/", 
                 ENCODE_FOR_URI(?name))) AS ?nameNode)
  BIND(IRI(CONCAT("https://ditrare.ise.fiz-karlsruhe.de/chemotion-kg/nodes/", 
                 ENCODE_FOR_URI(?url))) AS ?urlNode)
}
\end{lstlisting}

To ensure that the resulting ontology design patterns and their interconnections are transparent and reusable, a standardized graphical representation was adopted. The Chemotion-KG modules and their links were visualized using Graffoo, a formal graphical notation for OWL ontologies \cite{falco2014_graffoo_modelling}.

\section{Chemotion-KG Structure and Ontological Representation}
\label{sec:results}

The Chemotion-KG was constructed by transforming schema.org-based metadata into a BFO-aligned representation to ensure semantic interoperability with other scientific knowledge graphs. The choice of the Basic Formal Ontology (BFO) reflects its wide adoption as an upper ontology in the life sciences and research data management communities. Using BFO as a common top-level framework provides a consistent foundation for aligning domain-specific ontologies and supports integration with external resources that follow the same design principles. For modeling research data and its lifecycle, we adopted the NFDICore ontology, which is itself aligned with BFO. NFDICore was chosen because it already provides a well-defined vocabulary for metadata relevant to our setting—such as datasets, studies, creators, and affiliations—while also being developed and maintained within the German National Research Data Infrastructure (NFDI) context, ensuring institutional support and long-term sustainability. For representing chemical entities, we integrated the Chemical Entities of Biological Interest (ChEBI) ontology. ChEBI is a well-established community standard in chemistry, already aligned with BFO, and provides the necessary controlled vocabulary to represent molecules, substances, and their roles in experiments. Its wide acceptance in cheminformatics also facilitates linking the Chemotion-KG with other existing knowledge bases in chemistry.

\begin{figure}[ht]
    \centering
    \includegraphics[width=\textwidth]{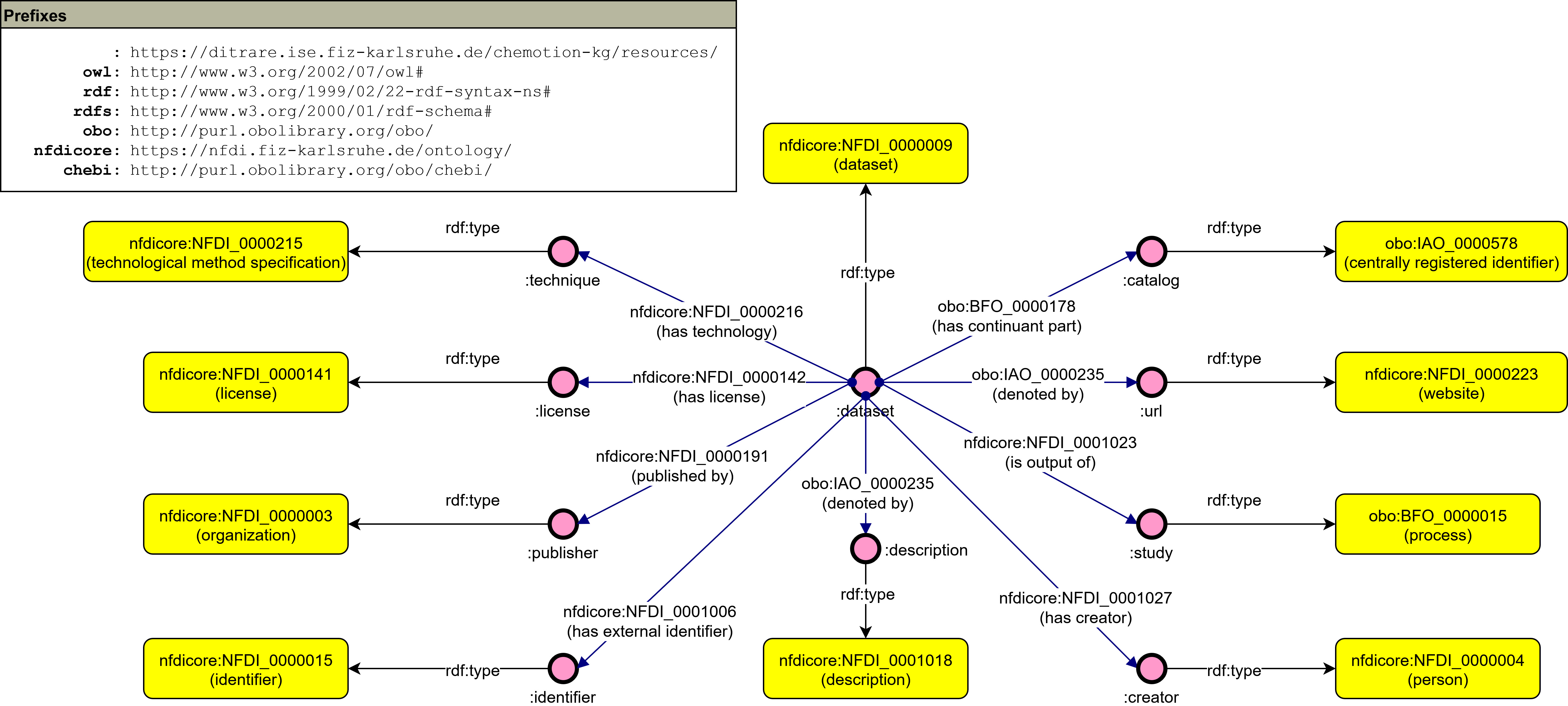}
    \caption{Dataset representation in the Chemotion-KG.}
    \label{fig:chemotion-dataset}
\end{figure}

Figure~\ref{fig:chemotion-dataset} illustrates the dataset representation, where schema.org \texttt{Dataset} instances were mapped to \texttt{nfdicore:NFDI\_0000009} and enriched with identifiers, licensing information, catalog registration, and links to associated studies. The class \texttt{nfdicore:NFDI\_0000009} represents a dataset, modeled as an information content entity that denotes a structured collection of data, typically organized for a defined purpose such as analysis, research, or reference. In this context, a dataset constitutes structured information about a resource curated or provided by an individual researcher. The alignment introduces explicit semantics for provenance and lifecycle stages through NFDICore and BFO relations. Provenance is explicitly captured via relations such as \texttt{nfdicore:NFDI\_0001027} (\emph{has creator}) and 
\texttt{nfdicore:NFDI\_0000191} (\emph{published by}), while lifecycle stages are represented using BFO continuant–occurrent 
relations like \texttt{obo:BFO\_0000178} (\emph{has continuant part}).

\begin{figure}[!t]
    \centering
    \includegraphics[width=\textwidth]{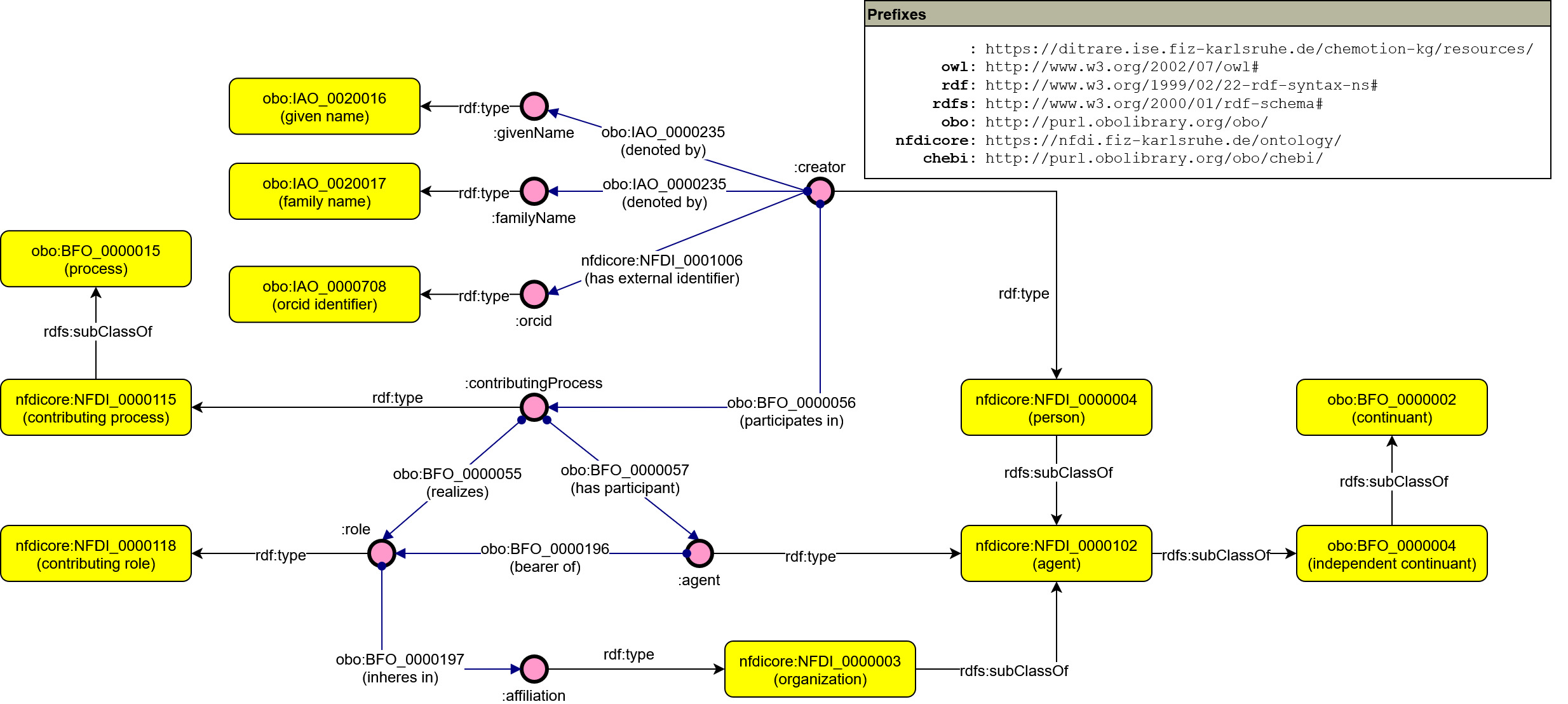}
    \caption{Creator description aligned to NFDICore and BFO patterns, enabling explicit representation of roles and affiliations.}
    \label{fig:chemotion-creator}
\end{figure}

The creator model (Figure~\ref{fig:chemotion-creator}) demonstrates the transformation of \texttt{schema:Person} metadata into a semantically rich structure. Individual names, ORCID identifiers, and organizational affiliations were represented as distinct resources, using \texttt{nfdicore:NFDI\_0000004} for persons and \texttt{nfdicore:NFDI\_0000003} for organizations. Roles and affiliations were explicitly modeled using the Process–Agent–Role ODP\footnote{\url{https://ise-fizkarlsruhe.github.io/mwo/docs/patterns/\#pattern-1-process-agent-role}}, aligning with the NFDICore and BFO modeling principles. This pattern establishes the relationship between \textit{bfo:Process} (Occurrent), \textit{nfdicore:Agent} (Independent Continuant), and \textit{bfo:Role} (Specifically Dependent Continuant). Within the Chemotion-KG, \textit{nfdicore:Agent} instances represent both organizations and persons involved in experimental chemistry workflows, including research institutions, and individual researchers. The pattern uses \texttt{bfo:has\_participant}, \texttt{bfo:realizes}, and \texttt{bfo:bearer\_of} to connect processes, agents, and their assigned roles, enabling explicit provenance tracking and role-based attribution for research data.

\begin{figure}[!t]
    \centering
    \includegraphics[width=0.8\textwidth]{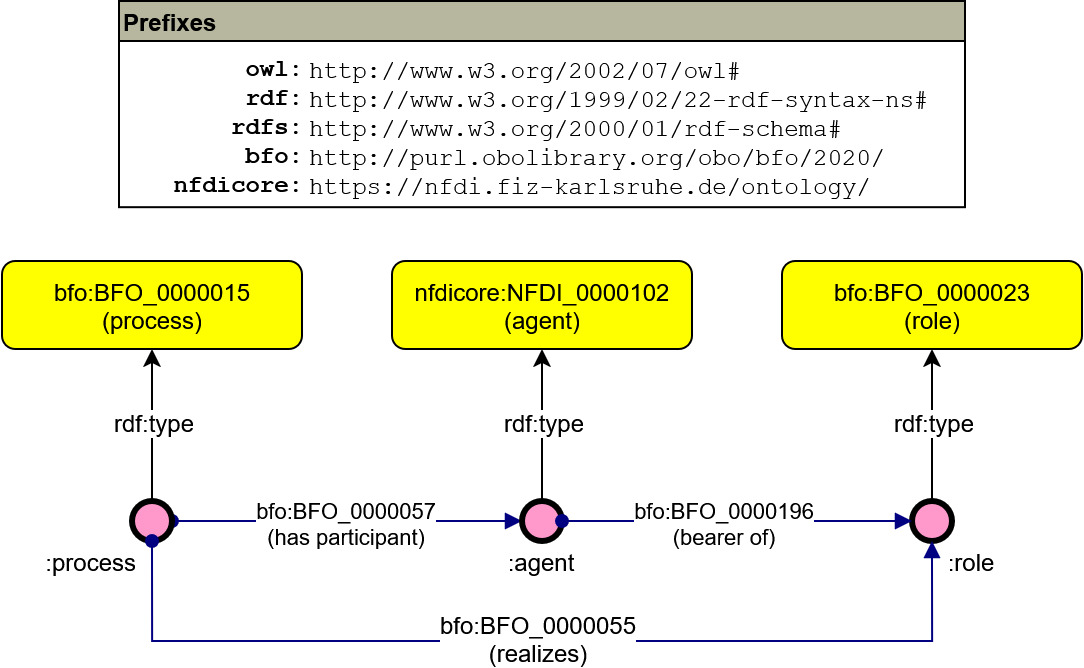}
    \caption{Process–Agent–Role Ontology Design Pattern.}
    \label{fig:process-agent-role-pattern}
\end{figure}

\begin{figure}[!t]
    \centering
    \includegraphics[width=\textwidth]{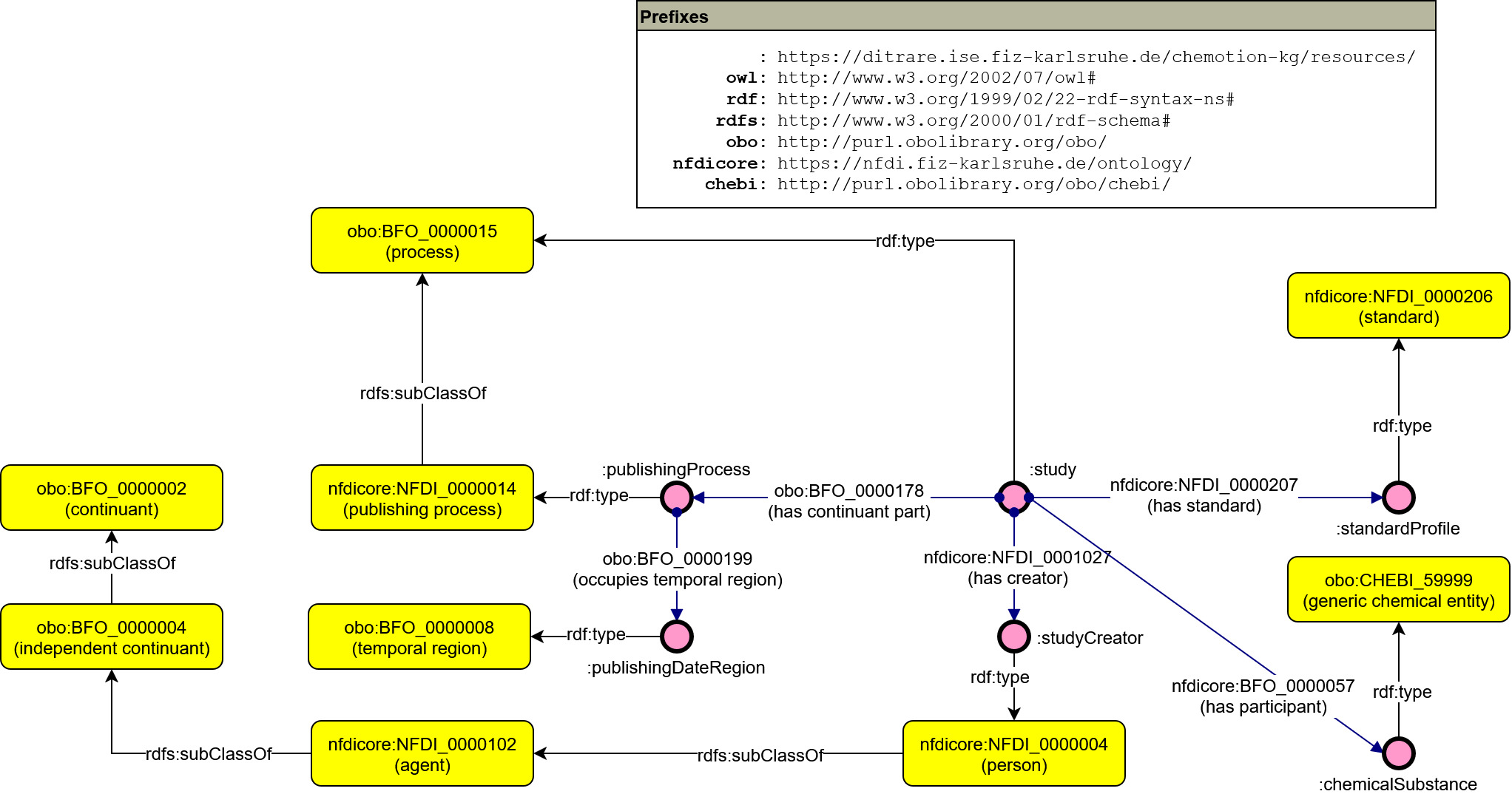}
    \caption{Study representation with explicit modeling of publishing processes, temporal regions, and standard profiles.}
    \label{fig:chemotion-study}
\end{figure}

\begin{figure}[!t]
    \centering
    \includegraphics[width=\textwidth]{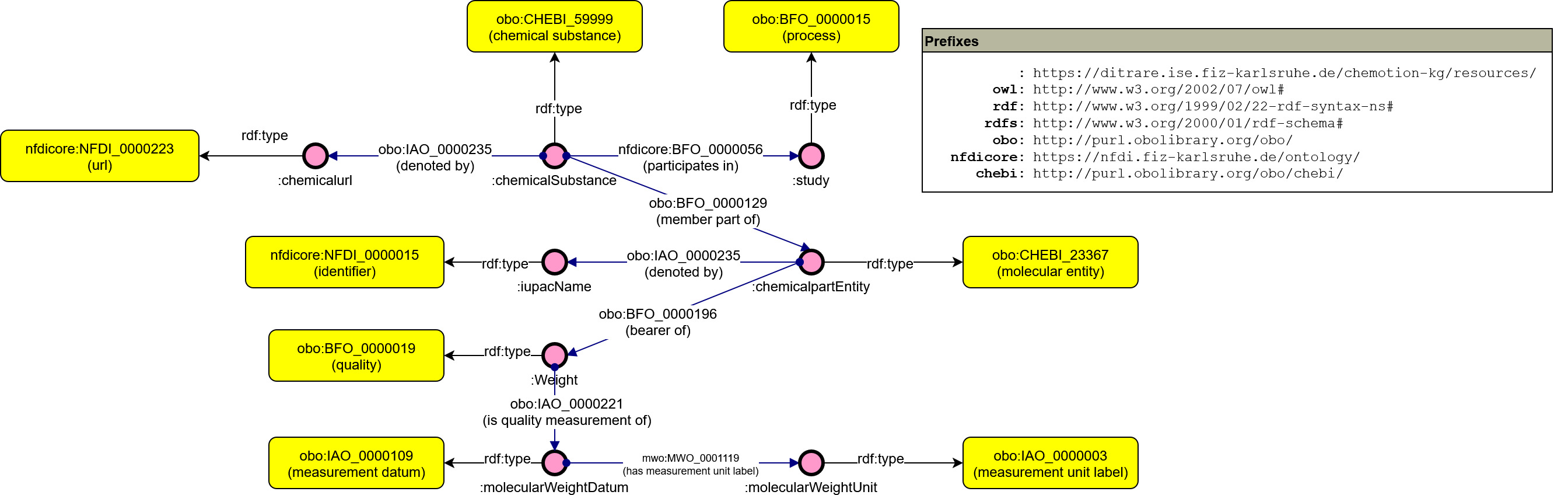}
    \caption{Chemical substance representation aligned to ChEBI and NFDICore, with explicit molecular entity and measurement modeling.}
    \label{fig:chemotion-chemicalSubstance}
\end{figure}

Studies were represented as \texttt{obo:BFO\_0000015} processes, connecting datasets, chemical substances, and associated publishing events 
(Figure~\ref{fig:chemotion-study}). The publishing activities were explicitly modeled using \texttt{nfdicore:NFDI\_0000014}, 
which is defined as a \emph{process of making information available to the public either for sale or free access}\footnote{\url{https://nfdi.fiz-karlsruhe.de/ontology/NFDI_0000014}}. 
Within the Chemotion-KG, this class was used to capture the temporal regions of publication via 
\texttt{obo:BFO\_0000199} (\emph{occupies temporal region}) and to link standard profiles with each study 
through \texttt{nfdicore:NFDI\_0000207} (\emph{has standard}).

Chemical entities (Figure~\ref{fig:chemotion-chemicalSubstance}) were aligned to ChEBI classes to ensure standardized representation of chemical knowledge. Generic chemical substances were modeled as \texttt{obo:CHEBI\_59999}, capturing the abstract notion of a chemical entity within a study, while specific molecular structures were represented as \texttt{obo:CHEBI\_23367} molecular entities.  To guarantee unambiguous identification and interoperability, the alignment included explicit mapping of InChI identifiers, InChIKeys, SMILES strings, and molecular formulas to structured nodes. Molecular weights were represented using the \texttt{obo:BFO\_0000019} \emph{quality} pattern combined with \texttt{obo:IAO\_0000109} \emph{measurement datum}, linking to \texttt{obo:IAO\_0000003} measurement units following the BFO measurement modeling principles. Structural images and external references were attached via \texttt{nfdicore:NFDI\_0000223} URL nodes to maintain provenance and facilitate linking to external chemical databases.

\begin{sidewaysfigure}
    \centering
    \includegraphics[width=\textheight,keepaspectratio]{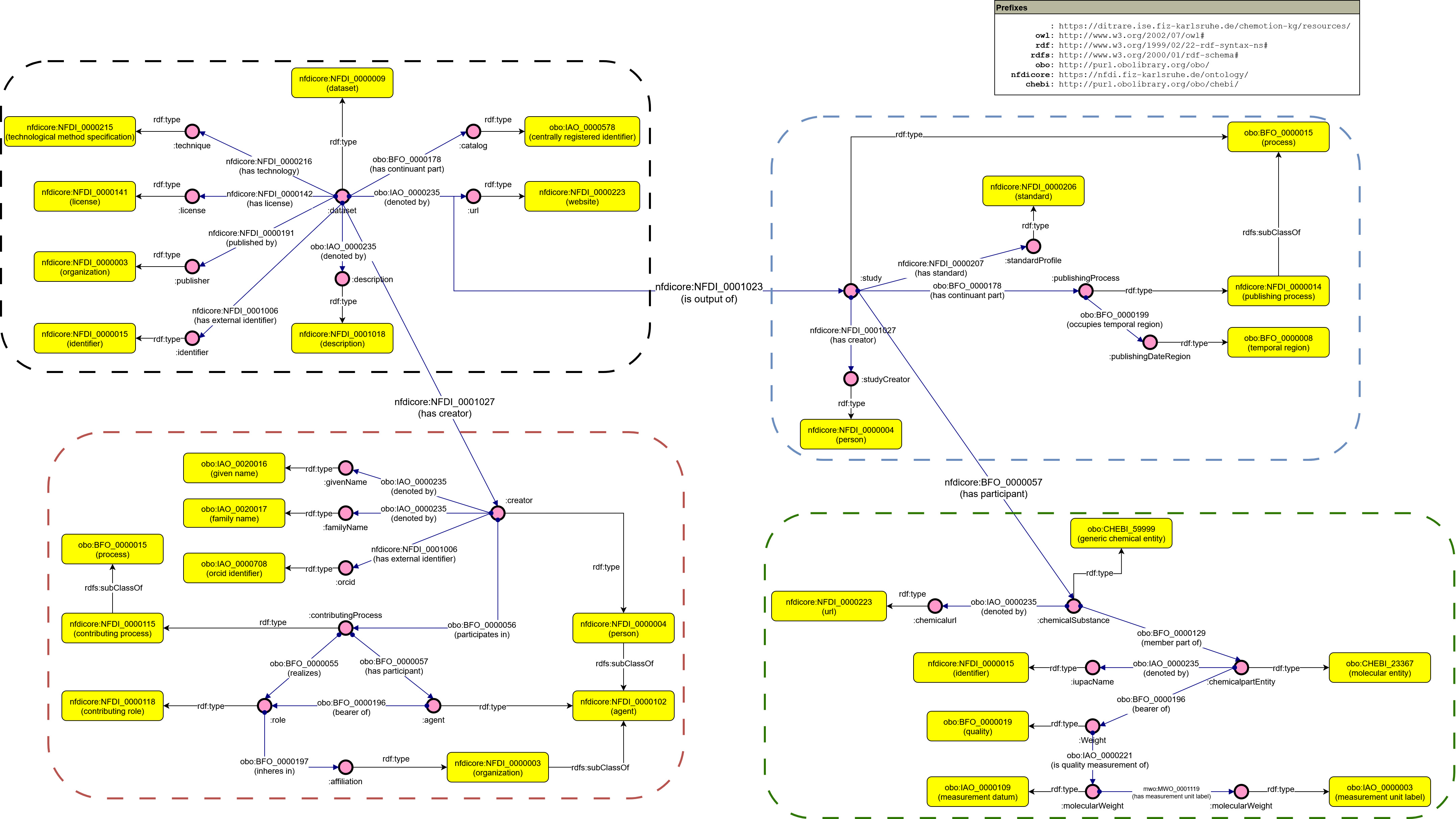}
    \caption{Integrated view of the Chemotion Knowledge Graph.}
    \label{fig:chemotion-all}
\end{sidewaysfigure}

The resulting Chemotion-KG provides a semantically enriched integration of experimental chemistry data, employing formal ODPs for each core entity type, including datasets, creators, studies, and chemical substances. The transformation pipeline applies a SPARQL CONSTRUCT-based approach that preserves the original \texttt{schema.org} metadata while aligning it with BFO semantics and enriching it through NFDICore and ChEBI ontologies. This results in a BFO-compliant graph that supports semantic interoperability, logical reasoning, and linkage to external knowledge resources. The overall Chemotion-KG integrates all ODPs into a coherent and interconnected model, capturing datasets, creators, studies, and chemical substances within a BFO-compliant framework. As illustrated in Figure~\ref{fig:chemotion-all}, each ODP is instantiated as a modular pattern and linked via well-defined object properties to form a semantically consistent network. This approach ensures that dataset descriptions, experimental studies, agents with roles, and chemical entities are harmonized through reusable modeling practices. The figure highlights how these ODPs are composed together to create a unified, semantically enriched representation of the Chemotion experimental data.

The Chemotion-KG is continuously updated, with new data from the Chemotion repository ingested into the graph on a daily basis. Instances are managed using a named graph that organizes resources according to their submission date. Each resource IRI encodes the year and month of submission alongside the original Chemotion identifier, ensuring temporal provenance and stable referencing. For example, the dataset representing Raman spectroscopy data is available at \footnote{\url{https://ditrare.ise.fiz-karlsruhe.de/chemotion-kg/resources/2014/05/10.14272/VRYFQVRFMNXTJS-UHFFFAOYSA-N/Raman}}, where the path segments \texttt{2014/05} capture the submission date and the suffix encodes the Chemotion-specific identifier. The corresponding source entry in the Chemotion repository\footnote{\url{https://www.chemotion-repository.net/inchikey/VRYFQVRFMNXTJS-UHFFFAOYSA-N/Raman}}, ensuring traceability between the knowledge graph and its originating records in the research data repository.

Access to the graph is provided through a public SPARQL endpoint at \url{https://ditrare.ise.fiz-karlsruhe.de/chemotion-kg/sparql}. This architecture combines automated metadata harvesting, ontology-based semantic enrichment, and fine-grained provenance modeling to create a high-quality, AI-ready knowledge graph for experimental chemistry. The constructed Chemotion-KG as of July 2025 comprises a total of 1,462,187 RDF triples, reflecting the semantic integration of experimental chemistry data and metadata. The graph contains 87,782 instantiated entities, including 20,701 datasets, 20,563 studies, and 3,746 molecular entities. The knowledge graph models 250 individual creators with explicit provenance, and integrates chemical information through 4,923 instances of obo:CHEBI\_59999 for generic chemical substances. 

The Chemotion-KG evolves continuously as new data are submitted to the Chemotion repository. Updates are handled through daily ingestion runs that add new instances and enrich existing ones, while preserving provenance by assigning each submission to a dedicated named graph organized by date. Entities are generally not removed; instead, updated resources are versioned through their temporal context in the named graph, ensuring reproducibility and historical traceability. IRIs are minted systematically by combining the Chemotion identifier with a submission timestamp, which mitigates the risk of collisions and guarantees long-term stability. To prevent excessive length, descriptive elements are restricted to standardized identifiers (e.g., InChIKey for molecules), while additional metadata are attached via well-defined properties rather than embedded directly in the IRI string.
IRIs in the Chemotion-KG are minted by combining the original Chemotion identifier with temporal submission information (year and month). This strategy provides globally unique and traceable references that preserve provenance across updates and reduces the likelihood of collisions. At present, descriptive identifiers (e.g., InChIKeys for molecules) are embedded in the IRI structure to maximize human interpretability and support external linking. However, this approach can result in relatively long IRIs, particularly for chemical description nodes, which may pose challenges for readability and downstream applications. To address this, we plan to adopt a Universally Unique Identifier (UUID) in future releases.

\clearpage
\clearpage

\section{Impact and Future Work}
\label{sec:impact}

The Chemotion-KG establishes a semantically enriched, BFO-compliant representation of experimental chemistry data derived from the Chemotion repository. By combining schema.org-based metadata with NFDICore and ChEBI ontologies through a SPARQL CONSTRUCT-driven pipeline, the approach ensures semantic interoperability, provenance preservation, and reasoning support. The daily ingestion workflow and ontology design patterns for datasets, creators, studies, and chemical substances provide a scalable foundation for AI-driven discovery and integration with external resources. The public SPARQL endpoint further facilitates community access, reuse, and advancement of FAIR chemical research data.

Future work will address two major directions: broadening the data scope and establishing systematic evaluation. On the data side, we will extend beyond metadata to incorporate additional information from the Chemotion repository model (e.g., raw instrument logs, measurement data, dataset details, and chemical substance information). We also plan to link the Chemotion-KG with external chemical knowledge graphs and databases, including PubChem RDF\footnote{\url{https://pubchem.ncbi.nlm.nih.gov/docs/rdf-federated-query}}\cite{Kim2025pubchem}, ChemSpider\footnote{\url{https://www.chemspider.com/}}\cite{Pence2010chemspider}, and the NFDI4Chem Knowledge Graph\footnote{\url{https://knowledgebase.nfdi4chem.de/knowledge_base/}}, thereby enhancing interoperability and enabling cross-resource reasoning. On the evaluation side, we will define competency questions that reflect typical use cases in chemistry, develop SHACL shapes to validate the RDF graphs generated in the Chemotion-KG pipeline, and conduct query performance benchmarks to assess scalability. In addition, we envision user studies with chemists to evaluate usability and scientific relevance. Finally, we will explore leveraging the Chemotion-KG for automated knowledge extraction and enrichment by integrating it with heterogeneous data sources and applying machine learning techniques for semantic alignment and knowledge discovery. In this context, we also plan to investigate the integration of large language models (LLMs) with structured scientific data to support advanced query answering.

The overarching goal of this work is to prepare the Chemotion system for integration with AI-based methods while embedding it into the broader ecosystem of semantically enriched research data management systems. In this context, knowledge graphs can be considered a component of symbolic AI, since they provide machine-readable semantics, enable logical entailments, and support query answering over structured knowledge. The Chemotion-KG contributes to this paradigm by supplying formally aligned, provenance-rich descriptions of experimental chemistry data that can serve as a foundation for both reasoning tasks and data-driven approaches. Example reasoning tasks include identifying equivalent substances across studies, detecting inconsistencies in dataset annotations, validating metadata through SHACL constraints, and inferring missing provenance relations. Beyond symbolic reasoning, we are exploring the use of AI methods such as natural language processing for automated curation of reaction descriptions and analysis modules within the Chemotion repository \cite{Punjabi2025nns,huang2024semi}. Linking Chemotion-KG to external chemical resources (e.g., PubChem, ChemSpider) will also enable cross-resource enrichment and the training of machine learning models for property prediction and reaction condition optimization. These efforts aim to bridge symbolic and statistical AI, ultimately supporting AI-assisted chemistry and laying the groundwork for self-driving laboratories \cite{Maffettone2023autondisc}.

Merging chemistry with AI has an immense potential for research. Studying this process and its effects is one of the foci of the Leibniz Science Campus DiTraRe. On a larger scale, DiTraRe brings an interdisciplinary perspective by investigating the influence of digitalisation of research. It sheds light on the potentials, challenges, and risks of the digital transformation through an interactive exchange between numerous research areas. Activities of the dimension ``Exploration and Knowledge Organisation'' in the Chemotion use case are strikingly similar to efforts with which researchers deal in other disciplines. More than the others, material science could serve here as a prime example, as it is already advancing the application of ontologies and knowledge graphs for experimental research \cite{Beygi2024matsc,norouzi2024landscape}. The corresponding German National Research Data Infrastructure, NFDIMatWerk, has already created a rich publication database concerning many topics which can be related to Chemotion-KG \cite{eberl_2024_nfdimatwerkproc}. DiTraRe plans to serve here as a platform connecting researchers from varying disciplines with an aim to exchange ideas, identify common problems, and create generalised solutions. This way, by raising above the level of individual research disciplines, we will study influence and effects of applied AI as an additional dimension to the digital transformation of research.

\section{Summary}
\label{sec:Summary}

In this paper, a comprehensive pipeline for constructing the BFO-compliant Chemotion Knowledge Graph has been presented, integrating experimental chemistry data from the Chemotion Repository into a semantically enriched, ontology-driven representation. By leveraging ontology design patterns and aligning schema-based metadata with NFDICore and ChEBI ontologies, the approach establishes a reusable and interoperable framework for chemical research data. The generated knowledge graph comprises over 1.4 million triples and more than 87 thousand instances.

\begin{acknowledgments}
  The Leibniz Science Campus ``Digital Transformation of Research'' (DiTraRe) is funded by the Leibniz Association (W74/2022). The authors would like to thank Pei-Chi Huang and Chia-Lin Lin (KIT IBCS) for their work on JSON-LD in Chemotion.
\end{acknowledgments}

%% The declaration on generative AI comes in effect
%% https://ceur-ws.org/GenAI/Policy.html
\section*{Declaration on Generative AI}
 \noindent During the preparation of this work, the author(s) used GPT-4 in order to: Grammar and spelling check. After using these tool(s)/service(s), the author(s) reviewed and edited the content as needed and take(s) full responsibility for the publication’s content. 
 
%% Define the bibliography file to be used
\bibliography{Norouzi+Chemotion-KG_Sci-K_2025}

\begin{thebibliography}{23}
\expandafter\ifx\csname natexlab\endcsname\relax\def\natexlab#1{#1}\fi
\providecommand{\url}[1]{\texttt{#1}}
\providecommand{\href}[2]{#2}
\providecommand{\path}[1]{#1}
\providecommand{\DOIprefix}{doi:}
\providecommand{\ArXivprefix}{arXiv:}
\providecommand{\URLprefix}{URL: }
\providecommand{\Pubmedprefix}{pmid:}
\providecommand{\doi}[1]{\href{http://dx.doi.org/#1}{\path{#1}}}
\providecommand{\Pubmed}[1]{\href{pmid:#1}{\path{#1}}}
\providecommand{\bibinfo}[2]{#2}
\ifx\xfnm\relax \def\xfnm[#1]{\unskip,\space#1}\fi
%Type = Article
\bibitem[{Steinbeck et~al.(2020)Steinbeck, Koepler, Bach, and et~al.}]{steinbeck2020nfdi4chem}
\bibinfo{author}{C.~Steinbeck}, \bibinfo{author}{O.~Koepler}, \bibinfo{author}{F.~Bach}, \bibinfo{author}{et~al.},
\newblock \bibinfo{title}{Nfdi4chem-towards a national research data infrastructure for chemistry in germany},
\newblock \bibinfo{journal}{Research ideas and outcomes} \bibinfo{volume}{6} (\bibinfo{year}{2020}) \bibinfo{pages}{e55852}.
%Type = Article
\bibitem[{Tremouilhac et~al.(2017)Tremouilhac, Nguyen, Huang, and et~al.}]{tremouilhac2017chemotion}
\bibinfo{author}{P.~Tremouilhac}, \bibinfo{author}{A.~Nguyen}, \bibinfo{author}{Y.-C. Huang}, \bibinfo{author}{et~al.},
\newblock \bibinfo{title}{Chemotion eln: an open source electronic lab notebook for chemists in academia},
\newblock \bibinfo{journal}{Journal of Cheminformatics} \bibinfo{volume}{9} (\bibinfo{year}{2017}) \bibinfo{pages}{54}.
%Type = Article
\bibitem[{Kotov et~al.(2018)Kotov, Tremouilhac, Jung, and et~al.}]{Kotov2018chemotion2}
\bibinfo{author}{S.~Kotov}, \bibinfo{author}{P.~Tremouilhac}, \bibinfo{author}{N.~Jung}, \bibinfo{author}{et~al.},
\newblock \bibinfo{title}{Chemotion-eln part 2: adaption of an embedded ketcher editor to advanced research applications.},
\newblock \bibinfo{journal}{Journal of Cheminformatics} \bibinfo{volume}{10} (\bibinfo{year}{2018}). \URLprefix \url{https://doi.org/10.1186/s13321-018-0292-9}. \DOIprefix\doi{10.1186/s13321-018-0292-9}.
%Type = Article
\bibitem[{Tremouilhac et~al.(2020)Tremouilhac, Lin, and Huang}]{Tremouilhac2020chemotionrepo}
\bibinfo{author}{P.~Tremouilhac}, \bibinfo{author}{C.-L. Lin}, \bibinfo{author}{P.-C. a.~a. Huang},
\newblock \bibinfo{title}{The repository chemotion: Infrastructure for sustainable research in chemistry},
\newblock \bibinfo{journal}{Angewandte Chemie International Edition} \bibinfo{volume}{59} (\bibinfo{year}{2020}) \bibinfo{pages}{22771--22778}. \URLprefix \url{https://onlinelibrary.wiley.com/doi/abs/10.1002/anie.202007702}. \DOIprefix\doi{https://doi.org/10.1002/anie.202007702}. \href{http://arxiv.org/abs/https://onlinelibrary.wiley.com/doi/pdf/10.1002/anie.202007702}{{\tt arXiv:https://onlinelibrary.wiley.com/doi/pdf/10.1002/anie.202007702}}.
%Type = Article
\bibitem[{Tremouilhac et~al.(2021)Tremouilhac, Huang, Lin, Huang, Nguyen, Jung, Bach, and Bräse}]{Tremouilhac2021chemotionrepo}
\bibinfo{author}{P.~Tremouilhac}, \bibinfo{author}{P.-C. Huang}, \bibinfo{author}{C.-L. Lin}, \bibinfo{author}{Y.-C. Huang}, \bibinfo{author}{A.~Nguyen}, \bibinfo{author}{N.~Jung}, \bibinfo{author}{F.~Bach}, \bibinfo{author}{S.~Bräse},
\newblock \bibinfo{title}{Chemotion repository, a curated repository for reaction information and analytical data},
\newblock \bibinfo{journal}{Chemistry methods} \bibinfo{volume}{1} (\bibinfo{year}{2021}) \bibinfo{pages}{8--11}. \DOIprefix\doi{10.1002/cmtd.202000034}.
%Type = Article
\bibitem[{Huang et~al.(2025)Huang, Lin, Tremouilhac, and et~al.}]{Huang2025chemotionrepo}
\bibinfo{author}{P.~Huang}, \bibinfo{author}{C.~Lin}, \bibinfo{author}{P.~Tremouilhac}, \bibinfo{author}{et~al.},
\newblock \bibinfo{title}{Using the chemotion repository to deposit and access fair research data for chemistry experiments},
\newblock \bibinfo{journal}{Nature Protocols} \bibinfo{volume}{20} (\bibinfo{year}{2025}). \URLprefix \url{https://doi.org/10.1038/s41596-024-01074-z}. \DOIprefix\doi{10.1038/s41596-024-01074-z}.
%Type = Article
\bibitem[{Huang et~al.(2021)Huang, Tremouilhac, Nguyen, and et~al.}]{Huang2021chemspectra}
\bibinfo{author}{Y.~Huang}, \bibinfo{author}{P.~Tremouilhac}, \bibinfo{author}{A.~Nguyen}, \bibinfo{author}{et~al.},
\newblock \bibinfo{title}{Chemspectra: a web-based spectra editor for analytical data},
\newblock \bibinfo{journal}{Journal of Cheminformatics} \bibinfo{volume}{18} (\bibinfo{year}{2021}). \URLprefix \url{https://doi.org/10.1186/s13321-020-00481-0}. \DOIprefix\doi{10.1186/s13321-020-00481-0}.
%Type = Incollection
\bibitem[{Gangemi and Presutti(2009)}]{gangemi2009ontologydesignpatter}
\bibinfo{author}{A.~Gangemi}, \bibinfo{author}{V.~Presutti},
\newblock \bibinfo{title}{Ontology design patterns},
\newblock in: \bibinfo{booktitle}{Handbook on ontologies}, \bibinfo{publisher}{Springer}, \bibinfo{year}{2009}, pp. \bibinfo{pages}{221--243}.
%Type = Misc
\bibitem[{Razum et~al.(2023)Razum, Bach, Brünger-Weilandt, and et~al.}]{ditrare_proposal}
\bibinfo{author}{M.~Razum}, \bibinfo{author}{F.~Bach}, \bibinfo{author}{S.~Brünger-Weilandt}, \bibinfo{author}{et~al.}, \bibinfo{title}{Proposal for a {L}eibniz {S}cience{C}ampus – {D}igital {T}ransformation of {R}esearch ({D}i{T}ra{R}e)}, \bibinfo{year}{2023}. \DOIprefix\doi{10.5281/zenodo.11109406}, \bibinfo{note}{project proposal}.
%Type = Misc
\bibitem[{Jacyszyn et~al.(2025)Jacyszyn, Bach, Sack, and et~al.}]{jacyszyn2025ditrare_interim}
\bibinfo{author}{A.~M. Jacyszyn}, \bibinfo{author}{F.~Bach}, \bibinfo{author}{H.~Sack}, \bibinfo{author}{et~al.}, \bibinfo{title}{Interim report of the leibniz science campus "digital transformation of research" (ditrare)}, \bibinfo{year}{2025}. \URLprefix \url{https://doi.org/10.5281/zenodo.14941635}. \DOIprefix\doi{10.5281/zenodo.14941635}.
%Type = Inproceedings
\bibitem[{Jacyszyn et~al.(2024)Jacyszyn, Sack, Group, and et~al.}]{jacyszyn2024ditrare}
\bibinfo{author}{A.~M. Jacyszyn}, \bibinfo{author}{H.~Sack}, \bibinfo{author}{D.-S. Group}, \bibinfo{author}{et~al.},
\newblock \bibinfo{title}{Ditrare: Ai on a spider’s web. interweaving disciplines for digitalisation},
\newblock in: \bibinfo{booktitle}{4th International Workshop on Scientific Knowledge: Representation, Discovery, and Assessment}, volume \bibinfo{volume}{Vol-3780}, \bibinfo{year}{2024}. \URLprefix \url{https://ceur-ws.org/Vol-3780/paper5.pdf}. \DOIprefix\doi{10.5281/zenodo.13862017}.
%Type = Book
\bibitem[{Arp et~al.(2015)Arp, Smith, and Spear}]{bfo2015}
\bibinfo{author}{R.~Arp}, \bibinfo{author}{B.~Smith}, \bibinfo{author}{A.~D. Spear}, \bibinfo{title}{{Building ontologies with Basic Formal Ontology}}, \bibinfo{publisher}{{MIT Press}}, \bibinfo{year}{2015}.
%Type = Article
\bibitem[{Bruns et~al.(2024)Bruns, Tietz, Waitelonis, and et~al.}]{bruns2024nfdicore}
\bibinfo{author}{O.~Bruns}, \bibinfo{author}{T.~Tietz}, \bibinfo{author}{J.~Waitelonis}, \bibinfo{author}{et~al.},
\newblock \bibinfo{title}{Nfdicore 2.0: A bfo-compliant ontology for multi-domain research infrastructures},
\newblock \bibinfo{journal}{arXiv preprint arXiv:2410.01821}  (\bibinfo{year}{2024}).
%Type = Article
\bibitem[{Del~Nostro et~al.(2022)Del~Nostro, Goldbeck, and Toti}]{CHAMEO}
\bibinfo{author}{P.~Del~Nostro}, \bibinfo{author}{G.~Goldbeck}, \bibinfo{author}{D.~Toti},
\newblock \bibinfo{title}{Chameo: An ontology for the harmonisation of materials characterisation methodologies},
\newblock \bibinfo{journal}{Applied Ontology}  (\bibinfo{year}{2022}) \bibinfo{pages}{1--21}.
%Type = Inproceedings
\bibitem[{Falco et~al.(2014)Falco, Gangemi, Peroni, and et~al.}]{falco2014_graffoo_modelling}
\bibinfo{author}{R.~Falco}, \bibinfo{author}{A.~Gangemi}, \bibinfo{author}{S.~Peroni}, \bibinfo{author}{et~al.},
\newblock \bibinfo{title}{Modelling owl ontologies with graffoo},
\newblock in: \bibinfo{booktitle}{European Semantic Web Conference}, \bibinfo{organization}{Springer}, \bibinfo{year}{2014}, pp. \bibinfo{pages}{320--325}.
%Type = Article
\bibitem[{Kim et~al.(2024)Kim, Chen, Cheng, and et~al.}]{Kim2025pubchem}
\bibinfo{author}{S.~Kim}, \bibinfo{author}{J.~Chen}, \bibinfo{author}{T.~Cheng}, \bibinfo{author}{et~al.},
\newblock \bibinfo{title}{Pubchem 2025 update},
\newblock \bibinfo{journal}{Nucleic Acids Research} \bibinfo{volume}{53} (\bibinfo{year}{2024}) \bibinfo{pages}{D1516--D1525}. \URLprefix \url{https://doi.org/10.1093/nar/gkae1059}. \DOIprefix\doi{10.1093/nar/gkae1059}. \href{http://arxiv.org/abs/https://academic.oup.com/nar/article-pdf/53/D1/D1516/60743708/gkae1059.pdf}{{\tt arXiv:https://academic.oup.com/nar/article-pdf/53/D1/D1516/60743708/gkae1059.pdf}}.
%Type = Article
\bibitem[{Pence and Williams(2010)}]{Pence2010chemspider}
\bibinfo{author}{H.~E. Pence}, \bibinfo{author}{A.~Williams},
\newblock \bibinfo{title}{Chemspider: An online chemical information resource},
\newblock \bibinfo{journal}{Journal of Chemical Education} \bibinfo{volume}{87} (\bibinfo{year}{2010}) \bibinfo{pages}{1123--1124}. \URLprefix \url{https://doi.org/10.1021/ed100697w}. \DOIprefix\doi{10.1021/ed100697w}. \href{http://arxiv.org/abs/https://doi.org/10.1021/ed100697w}{{\tt arXiv:https://doi.org/10.1021/ed100697w}}.
%Type = Article
\bibitem[{Punjabi et~al.(2025)Punjabi, Huang, Holzhauer, and et~al.}]{Punjabi2025nns}
\bibinfo{author}{D.~Punjabi}, \bibinfo{author}{Y.~Huang}, \bibinfo{author}{L.~Holzhauer}, \bibinfo{author}{et~al.},
\newblock \bibinfo{title}{Infrared spectrum analysis of organic molecules with neural networks using standard reference data sets in combination with real-world data},
\newblock \bibinfo{journal}{Journal of Cheminformatics} \bibinfo{volume}{17} (\bibinfo{year}{2025}). \URLprefix \url{https://doi.org/10.1186/s13321-025-00960-2}. \DOIprefix\doi{10.1186/s13321-025-00960-2}.
%Type = Article
\bibitem[{Huang et~al.(2024)Huang, Tremouilhac, Kuhn, and et~al.}]{huang2024semi}
\bibinfo{author}{Y.-C. Huang}, \bibinfo{author}{P.~Tremouilhac}, \bibinfo{author}{S.~Kuhn}, \bibinfo{author}{et~al.},
\newblock \bibinfo{title}{(semi-) automatic review process for common compound characterization data in organic synthesis},
\newblock \bibinfo{journal}{ChemRxiv}  (\bibinfo{year}{2024}). \DOIprefix\doi{10.26434/chemrxiv-2024-1r9tb}, \bibinfo{note}{preprint}.
%Type = Article
\bibitem[{Maffettone et~al.(2023)Maffettone, Friederich, Baird, and et~al.}]{Maffettone2023autondisc}
\bibinfo{author}{P.~M. Maffettone}, \bibinfo{author}{P.~Friederich}, \bibinfo{author}{S.~G. Baird}, \bibinfo{author}{et~al.},
\newblock \bibinfo{title}{What is missing in autonomous discovery: open challenges for the community},
\newblock \bibinfo{journal}{Digital Discovery} \bibinfo{volume}{2} (\bibinfo{year}{2023}) \bibinfo{pages}{1644--1659}. \URLprefix \url{http://dx.doi.org/10.1039/D3DD00143A}. \DOIprefix\doi{10.1039/D3DD00143A}.
%Type = Article
\bibitem[{Beygi~Nasrabadi et~al.(????)Beygi~Nasrabadi, Norouzi, Sack, and Skrotzki}]{Beygi2024matsc}
\bibinfo{author}{H.~Beygi~Nasrabadi}, \bibinfo{author}{E.~Norouzi}, \bibinfo{author}{H.~Sack}, \bibinfo{author}{B.~Skrotzki},
\newblock \bibinfo{title}{Performance evaluation of upper-level ontologies in developing materials science ontologies and knowledge graphs},
\newblock \bibinfo{journal}{Advanced Engineering Materials} \bibinfo{volume}{n/a} (????) \bibinfo{pages}{2401534}. \URLprefix \url{https://onlinelibrary.wiley.com/doi/abs/10.1002/adem.202401534}. \DOIprefix\doi{https://doi.org/10.1002/adem.202401534}. \href{http://arxiv.org/abs/https://onlinelibrary.wiley.com/doi/pdf/10.1002/adem.202401534}{{\tt arXiv:https://onlinelibrary.wiley.com/doi/pdf/10.1002/adem.202401534}}.
%Type = Inproceedings
\bibitem[{Norouzi et~al.(2024)Norouzi, Waitelonis, and Sack}]{norouzi2024landscape}
\bibinfo{author}{E.~Norouzi}, \bibinfo{author}{J.~Waitelonis}, \bibinfo{author}{H.~Sack},
\newblock \bibinfo{title}{The landscape of ontologies in materials science and engineering: a survey and evaluation},
\newblock in: \bibinfo{booktitle}{Proceedings of the First International Workshop on Semantic Materials Science (SeMatS 2024): Harnessing the Power of Semantic Web Technologies in Materials Science}, volume \bibinfo{volume}{3760} of \textit{\bibinfo{series}{CEUR Workshop Proceedings}}, \bibinfo{publisher}{CEUR-WS.org}, \bibinfo{year}{2024}, pp. \bibinfo{pages}{78--100}. \URLprefix \url{https://ceur-ws.org/Vol-3760/paper3.pdf}.
%Type = Proceedings
\bibitem[{ebe(2024)}]{eberl_2024_nfdimatwerkproc}
\bibinfo{title}{Proceedings NFDI-MatWerk Conference 2023}, \bibinfo{publisher}{Zenodo}, \bibinfo{year}{2024}. \URLprefix \url{https://doi.org/10.5281/zenodo.11353339}. \DOIprefix\doi{10.5281/zenodo.11353339}.

\end{thebibliography}

%\appendix

%\section{Appendix section}

\end{document}